\documentclass[a4paper, 11pt]{article}

\title{On click-fraud under pro-rata revenue sharing rule\thanks{
The author thanks Anna Bogomolnaia, Hervé Moulin, Xiaochang Lei, Georgios Gerasimou, Yiannis Vailakis, and an anonymous referee for helpful suggestions, as well as audiences of Microtheory internal seminar at University of Glasgow. All remaining errors are my own.
}
}

\author{Hao Yu\thanks{
Adam Smith Business School, University of Glasgow. Email: \href{mailto:hao.yu@glasgow.ac.uk}{hao.yu@glasgow.ac.uk}
}
}

\date{\today}
\usepackage[margin=1.25in]{geometry}

\usepackage{amsmath, amsfonts, amsthm, amssymb, graphicx, enumitem}
\usepackage[colorlinks=true,citecolor=blue]{hyperref}
\usepackage[round]{natbib}

\usepackage{fontenc}[T1]
\usepackage{libertine}

\newtheorem{definition}{Definition}
\newtheorem{theorem}{Theorem}
\newtheorem{proposition}{Proposition}
\newtheorem{corollary}{Corollary}
\newtheorem{lemma}{Lemma}

\selectfont
\begin{document}

\maketitle

\begin{abstract}
    Click-fraud is commonly seen as a key vulnerability of pro-rata revenue sharing rule on music streaming platforms, whereas user-centric is largely immune. This paper develops a tractable non-cooperative model in which artists can purchase fraud activity that generates undetectable fake streams up to a technological limit. We defend pro-rata by showing that it is fraud-robust: when fraud technology is weak, honesty is a strictly dominant strategy, and an efficient fraud-free equilibrium obtains; when fraud technology is strong, a unique fraud equilibrium arises, yet aggregate fake streams remain bounded. Although fraud is inefficient, the resulting redistribution may improve fairness in some cases. To mitigate fraud without abandoning pro-rata, we introduce a parametric weighted rule that interpolates between pro-rata and user-centric, and characterize parameter ranges that restore a fraud-free equilibrium under technology constraint. We also discuss implications of Spotify's modernized royalty system for fraud incentives.
    \\[0.5cm]
\end{abstract}

\newpage

\section{Introduction}
Music streaming platforms (e.g., Spotify, Apple Music, and Deezer) charge users a subscription fee and share revenue with artists who supply music content. Royalties are typically allocated as a function of users' streams and the platform's revenue-sharing rule. Two rules commonly used in practice are pro-rata and user-centric. Under pro-rata, an artist receives a share equal to their streams divided by total streams; under user-centric, each user's payment is allocated proportionally to the artists that user streams.

There is an ongoing debate about which rule should be used, especially between pro-rata and user-centric. At the center of this debate, \cite{dimont2018} raises several major concerns about pro-rata, and advocates a pay-per-subscriber model, such as user-centric.\footnote{In addition to the problem of disproportionate royalties and click-fraud mentioned in this paper, \cite{dimont2018} also raises a concern about the silence of major labels. We omit this here because it is unrelated to the pro-rata mechanism itself.} One concern is the problem of disproportionate royalties. The argument highlights cross-subsidization from low-usage users to high-usage users, as well as the induced cross-subsidization between artists listened to by heavy users and those listened to by light users. \cite{alaei2022} defend pro-rata on this point, showing that despite cross-subsidization across users, pro-rata can be preferred by both artists and the platform.

Another major concern is click-fraud. The intuition is straightforward: if an artist can generate an unlimited number of artificial streams under pro-rata, they could capture nearly all royalties. Real-world examples include Vulfpeck's Sleepify and Michael Smith's bot accounts, as discussed in the literature \citep{dimont2018, ghosh2025}. Vulfpeck created an album consisting of silent tracks and asked fans to stream it while asleep. Michael Smith, who was ultimately arrested, uploaded AI-generated songs and used thousands of bot accounts to boost streams. In both cases, fraudulent activity increased royalty payments.

Strategic manipulation by artists, i.e., click-fraud, has been studied from an axiomatic perspective. \citet{bergantinos2025b} propose an axiom called click-fraud proofness, motivated by Vulfpeck's Sleepify. It requires that an artist cannot gain more than the subscription fee when a user reallocates streams. \citet{ghosh2025} strengthen this requirement with single-user bribery-proofness, which bounds the bonus to any subset of artists by the subscription fee paid by an additional user. The single-user fraud-proofness axiom in \citet{ghosh2025} captures the logic behind Michael Smith's bot accounts: an artist should not profit from faking a new user by more than the subscription cost. User-centric satisfies these axioms, whereas pro-rata does not.

However, this does not mean that pro-rata's vulnerability to fraud is insurmountable. As noted in \citet{ghosh2025}, these axioms impose no upper bound on streams generated by manipulated users. In practice, even without sophisticated fraud detection, platforms can detect and remove large volumes of suspicious streams. To capture this idea, we model a fraud technology that produces a fixed number of streams, $\lambda_0$, per unit of fraud activity purchased by artists. The parameter $\lambda_0$ represents the maximum number of streams from each fake user that the platform fails to detect. Interpreted in relative terms, improved detection reduces the scope for subtle manipulation, while evolving fraud methods may make manipulation harder to detect.

This paper defends pro-rata with respect to click-fraud. Previewing the main results, we show that pro-rata is fraud-robust: honesty is a strictly dominant strategy for all artists when fraud technology is not too strong, with a threshold that depends on artists' real streamshares. Even when fraud technology is strong and cannot be fully controlled, total fake streams remain bounded. Under pro-rata, an increase in total streams dilutes the value of each stream, reducing the marginal benefit of additional fake streams until it falls below the fraud cost. Moreover, fraud may lead to a fairer distribution: it benefits cheaters while reducing the income of high-streamshare honest artists by diluting stream value, potentially increasing the minimum payoff. Thus, the problem of disproportionate royalties may also be mitigated under pro-rata due to click-fraud.

We also discuss alternative approaches. First, we extend the analysis to a class of weighted rules that are convex combinations of pro-rata and user-centric. These rules differ in their ability to deter fraud and allow the platform to balance simplicity and robustness while maintaining efficiency under constraints on fraud technology. Second, we discuss Spotify's recent modernization of its royalty system, focusing on a qualification policy that introduces an eligibility threshold for revenue sharing. We show that when the qualification threshold is not too high, the policy either fails to deter fraud or harms all artists.

This paper is the first to study strategic manipulation by artists on music streaming platforms using a non-cooperative game. By analyzing pro-rata's vulnerability to fraud, the paper contributes to the ongoing debate between pro-rata and user-centric. Beyond conceptual discussions and axiomatic analyses \citep{dimont2018, bergantinos2025b, ghosh2025}, \citet{alaei2022} and \citet{lei2023} are the main theoretical contributions in this area. \citet{alaei2022} study a model with participation decisions by both users and artists. \citet{lei2023} shows that pro-rata can be more efficient and fairer than user-centric from users' perspectives when music quality is endogenously chosen. In contrast to these long-run approaches that emphasize participation and quality, our paper focuses on artists' short-run strategic manipulation, thereby enriching this literature.

Contest technology is applied in the analysis. Fraud activity links the pro-rata revenue-sharing rule to standard contest models with head starts. The contest success function in our model is axiomatized by \citet{yu2025} through ``No advantageous reallocation,'' extending \citet{amegashie2006} from the symmetric case to a contest framework with equal starts. \citet{franke2018} study lottery contests with linear contest technology in a setting more general than ours, focusing on the optimal design of headstarts to maximize effort. In our model, efforts are purely socially wasteful, and the focus is on equilibrium conditions, bounds on aggregate fraud, and fairness comparisons.

The remainder of the paper is organized as follows. Section \ref{sec:model} presents the model. Section \ref{sec:case1} characterizes the fraud-free equilibrium under weak fraud technology. Section \ref{sec:case2} analyzes the fraud equilibrium under strong fraud technology. Section \ref{sec:fairness} compares the fairness driven by fraud. Section \ref{sec:weighted} discusses alternative solutions. Section \ref{sec:conclusion} concludes.

\section{Model}\label{sec:model}
Consider a two-sided music platform that hosts both users and artists. Each user pays a subscription fee normalized to 1, and artists share the $\beta\in(0,1)$ fraction of total revenue. Denote by $N$ the set of $n$ artists and by $M$ the set of $m$ users. For each user $j\in M$, let the total number of streams be $\lambda_j>0$, and let $\pi_{ij}\in[0,1]$ denote the share of user $j$'s streams allocated to artist $i\in N$, with $\sum_{i\in N}\pi_{ij}=1$. Thus, the streams from user $j$ on artist $i$ are $\lambda_j\pi_{ij}$. The model is considered over a short subscription term (e.g., monthly). Over the short term, artist participation is fixed. Combining full participation of artists with constant user preferences, the set of users maintaining their subscriptions is also unchanged \citep{alaei2022}. Thus $\lambda_j$ and $\pi_{ij}$ are constant for all $i\in N$ and $j\in M$.

Revenue-sharing rules allocate the revenue $m\beta$ to artists based on users' streams. Under pro-rata, revenue is shared among artists according to their streamshares. The revenue paid to artist $i$ is
\begin{equation*}
 R_i^{pr} = \frac{\sum_{j\in M}\lambda_j \pi_{ij}}{\sum_{j\in M}\lambda_j }m\beta.
\end{equation*}
In contrast, user-centric allocates each user's contribution $\beta$ across artists in proportion to that user's streaming shares. It gives artist $i$ the revenue
\begin{equation*}
 R_i^{uc} = \sum_{j\in M}\pi_{ij} \beta.
\end{equation*}
Let $\overline{\lambda} \equiv \frac{1}{m}\sum_{j\in M}\lambda_j$ denote the average number of streams per user, and $d_i\equiv\frac{\sum_{j\in M}\lambda_j\pi_{ij}}{\sum_{j\in M}\lambda_j}$ denote artist $i$'s share of total streams. Then the pro-rata payment can be written as $R_i^{pr}=d_i m\beta$. Assume $d_i<1$ for all $i\in N$ to exclude the trivial case of a monopolist artist.

The model is a non-cooperative game in which the only strategic agents are artists. Each artist $i\in N$ chooses a level of fraud activity $x_i\ge 0$, and let $x\in\mathbb{R}_+^N$ denote the strategy profile. Fraud activity has unit cost $\delta>1$, which covers the subscription fee. Each unit of fraud activity creates a new user who generates $\lambda_0$ fake streams on the cheating artist. The parameter $\lambda_0$ represents the maximum number of streams per fake user that the platform cannot detect or remove; we refer to $\lambda_0$ as the fraud technology, as it captures the platform's ability to detect fake streams.

Given $x$, the net gain of artist $i$ under pro-rata is
\begin{equation*}
 G_i^{pr}(x) = \frac{m \overline{\lambda} d_i + \lambda_0 x_i}{m\overline{\lambda} + \lambda_0\sum_{j\in N}x_j} \Bigg(m + \sum_{j\in N}x_j\Bigg)\beta - \delta x_i, \label{eq:ind1}
\end{equation*}
where the fraction is artist $i$'s revenue share, $(m+\sum_{j\in N}x_j)\beta$ is the total revenue allocated to artists, and $\delta x_i$ is the fraud cost. In the numerator, $m\overline{\lambda}d_i=\sum_{k\in M}\lambda_k\pi_{ik}$ denotes real streams on artist $i$, and $\lambda_0 x_i$ denotes fake streams generated by artist $i$. In the denominator, $m\overline{\lambda}=\sum_{k\in M}\lambda_k$ is total real streams, and $\lambda_0\sum_{j\in N}x_j$ is total fake streams.

Remark that user-centric is immune to fraud \citep{bergantinos2025b, ghosh2025}. Under user-centric, the net gain of artist $i$ is
\begin{equation*}
 G_i^{uc}(x) = \Bigg(\sum_{j\in M}\pi_{ij} + x_i\Bigg)\beta - \delta x_i,
\end{equation*}
where a unit of fraud activity increases the cheating artist's revenue by $\beta$ while costing $\delta$. That is, fraud is not profitable under user-centric.

To simplify the analysis, let $\xi=\frac{\delta-\beta}{\beta}$ and $V=\frac{\lambda_0-\overline{\lambda}}{\xi\overline{\lambda}}$. Define the ratio of fake streams on artist $i$ to total real streams $t_i=\frac{\lambda_0 x_i}{m\overline{\lambda}}$ for all $i\in N$. In the simplified model, each artist $i$ choose strategy $t_i\in \mathbb{R}_+$. Define the utility of artist $i\in N$ as
$u_i(t)=\frac{\lambda_0 G_i^{pr}(x)}{m\overline{\lambda}(\delta-\beta)}$,
which yields
\begin{equation}
 u_i(t) = \frac{d_i + t_i}{1 + \sum_{j\in N}t_j}V - t_i + \frac{d_i}{\xi}.
\end{equation}
The last term is constant and therefore does not affect incentives. All information in the model is public. In the following analysis, we focus on dominant strategy equilibrium when the fraud technology is weak, and on Nash equilibrium otherwise.

\section{Case 1: Weak fraud technology}\label{sec:case1}

In this paper, efficiency is measured by the total surplus of artists under a utilitarian criterion. We call an equilibrium \textit{efficient} if it maximizes the summation of artists' utilities. Note that each unit of fraud activity adds a taxed subscription fee $\beta$ to the revenue pool shared by artists, but incurs a fraud cost $\delta$ for the cheating artist. 
Thus, fraud is socially wasteful: it reduces total surplus by $\delta-\beta>0$ per unit of fraud activity. Consequently, the most desirable outcome would be an equilibrium in which all artists remain honest, thereby avoiding social waste.

To illustrate the possibility of this desirable outcome, we first show that honesty is a strictly dominant strategy when fraud technology is weak.

\begin{lemma} \label{lem:index2}
 $t_i = 0$ is a strictly dominant strategy of artist $i$ if and only if $\lambda_0 \leq \overline{\lambda}[1+\xi h(d_i)]$,
 where
 \begin{equation*}
 h(d_i) = \left\{\begin{array}{ll}
 \frac{1}{1-d_i} & \mathrm{if}\ d_i\in[0, \frac{1}{2}]; \\
 4d_i & \mathrm{if}\ d_i \in (\frac{1}{2}, 1].
 \end{array}\right.
 \end{equation*}
\end{lemma}

The function $h(\cdot)$ is continuous and strictly increasing. It implies that the decision to engage in fraud depends on the artist's real streamshare $d_i$. An equivalent condition for honesty can be written as
\begin{equation*}
 \frac{\lambda_0 - \overline{\lambda}}{\overline{\lambda}} \leq \xi h(d_i).
\end{equation*}
From this expression, $\xi h(\cdot)$ represents a threshold for the deviation of fraud technology from average streaming. Intuitively, the left-hand side (LHS) measures how closely a fake user resembles a real user: the smaller the deviation, the more subtle the manipulation. If the platform can detect and remove all fake users whose streaming is no greater than that of real users (i.e., LHS $\le 0$), then fraud is not profitable and all artists remain honest.

We interpret $\xi$ as the fraud premium and decompose it as
\begin{equation*}
 \xi = \frac{1}{\beta}[(\delta-1) + (1-\beta)],
\end{equation*}
where $\delta-1$ is the price of the fraud service and $1-\beta$ is the platform's retained share of the subscription fee. The fraud premium measures the extra cost per fake user relative to the maximum surplus $\beta$ obtainable from fraud through the revenue pool.

We refer to an equilibrium $t^*$ as a fraud-free equilibrium if there is no fake stream generated, i.e., $t^*=\mathbf{0}$.
As an immediate implication of Lemma \ref{lem:index2}, the following theorem characterizes the fraud-free equilibrium.

\begin{theorem}\label{thm:index2}
    $t^* = \mathbf{0}$ is the dominant strategy equilibrium if and only if the fraud technology satisfies
 \begin{equation}
     \lambda_0 \leq \overline{\lambda}\bigg(1 + \frac{\xi}{1-d_{\min}}\bigg),\label{eq:ind2}
 \end{equation}
 where $d_{\min}\equiv\min_{i\in N}d_i$.
\end{theorem}

We call the fraud technology \textit{weak} if \eqref{eq:ind2} holds, and \textit{strong} otherwise.
Theorem \ref{thm:index2} shows that fraud can be deterred under a sufficiently weak fraud technology. Hence a unique fraud-free equilibrium exists. 
This fraud-free equilibrium is efficient because it involves no waste. 
In this equilibrium, honesty is a strictly dominant strategy for each artist, aligning the incentive with the utilitarian objective.

Theorem \ref{thm:index2} also suggests two levers for mitigating fraud. First, the platform can adjust the revenue share parameter $\beta$; in the long run, however, such changes may affect participation \citep{alaei2022}. Second, the platform can strengthen quality checks, e.g., identity verification, to increase the cost of fraud services $\delta$ and hence the premium $\xi$.
The threshold is increasing in the pivotal artist's streamshare $d_{\min}$. However, balancing streamshares among artists can raise the threshold only up to $\overline{\lambda}(1+\frac{n}{n-1}\xi)$, which may be limited in a large platform ($n\to+\infty$).

\section{Case 2: Strong fraud technology}\label{sec:case2}
When fraud technology is sufficiently strong, the efficient fraud-free equilibrium collapses, and cheating emerges as a rational strategy.

\begin{lemma}\label{lem:index12}
 If the fraud technology is sufficiently strong, i.e., $\lambda_0 > \overline{\lambda}(1 + \frac{\xi}{1-d_{\min}})$, then there exists a unique pure strategy Nash equilibrium. This equilibrium is a fraud equilibrium.
\end{lemma}

The fraud equilibrium refers to an equilibrium with positive fraud activity. Once fraud technology exceeds the critical level, honesty is no longer sustainable. In particular, the pivotal artist with the lowest real streamshare $d_{\min}$ has an incentive to cheat.

Theorem \ref{thm:index4} characterizes which artists cheat and by how much.

\begin{theorem}\label{thm:index4}
 In the fraud equilibrium $t^*$, there exists a unique non-empty set of dishonest artists $N^d\subseteq N$ with cardinality $n_d$, such that $t_i^* > 0$ for all $i\in N^d$ and $t_j^* = 0$ for all $j\in N\setminus N^d$. The dishonest artists are those with low real streamshares.

 \begin{enumerate}[label=(\alph*)]
 \item For all dishonest artists $i,j\in N^d$ with $t_i^*, t_j^* > 0$, we have $d_i + t_i^* = d_j + t_j^*$.
 \item Let $T^* = \sum_{i\in N^d}t_i^*$. The ratio of total streams to real streams is
 \begin{equation}
 1 + T^* = \sqrt{\bigg(\frac{n_d-1}{2n_d}V\bigg)^2 + \frac{1}{n_d}\Bigg(1-\sum_{i\in N^d}d_i\Bigg)V} + \frac{n_d-1}{2n_d}V.\label{eq:index32}
 \end{equation}
 \end{enumerate}
\end{theorem}

Next consider the upper bound on the ratio of total streams to real streams $1+T^*$.

\begin{proposition}\label{prop:index1}
 The upper bound of the total-to-real streams ratio $1+T^*$ is $V$.
\end{proposition}

Recall that $\frac{\lambda_0 - \overline{\lambda}}{\overline{\lambda}}$ measures the relative advantage obtained from fraud, and $\xi$ is the fraud premium, measuring the percentage extra cost. Thus, $V = \frac{\lambda_0 - \overline{\lambda}}{\xi \overline{\lambda}}$ summarizes the relative advantage from socially wasteful manipulation.

This result is reassuring because even under extreme conditions, total fraud cannot diverge without bound. It suggests that axioms based on unlimited strategic manipulation may be overly strong in this setting \citep{bergantinos2025b, ghosh2025}: even the summation of all artists' fraud activities is bounded.

Because dishonest artists are those with low real streamshares, the case in which all artists engage in fraud is referred to as the \textit{worst fraud equilibrium}. In this case, $N^d=N$ and $n_d=n$. The ratio of total streams to real streams in the worst fraud equilibrium is $\frac{n-1}{n}V$, since $\sum_{i\in N}d_i=1$. In a large platform ($n\to+\infty$), the ratio of fake streams to real streams approaches the maximum $V-1$.

The following corollary provides an equivalent condition for when the worst case occurs.

\begin{corollary}\label{coro:index1}
 The fraud equilibrium is the worst if and only if
 \begin{equation*}
     d_{\mathrm{max}} < \frac{n-1}{n^2}V,
 \end{equation*}
 where $d_{\mathrm{max}}\equiv \max_{i\in N}d_i$.
\end{corollary}

This corollary links the largest real streamshare to the worst-case condition: the artist with the largest real streamshare is the least likely to cheat. If $d_{\mathrm{max}}$ is bounded away from zero and entry is free in a large platform, the worst case will not occur.

Now consider a \textit{partial fraud equilibrium}, where not all artists engage in fraud. By Theorem \ref{thm:index4}, cheating occurs among low-streamshare artists, and equilibrium fraud is decreasing in real streamshare. Hence there exists a cutoff $d^*$ such that artists with $d_i<d^*$ belong to $N^d$, while those with $d_i\ge d^*$ remain honest.

\begin{corollary}\label{coro:index2}
 There exists a cutoff $d^*$ distinguishing honest artists $j\in N\setminus N^d$ with $d_j \geq d^*$ from dishonest artists $i\in N^d$ with $d_i < d^*$ in a partial fraud equilibrium. That is,
 \begin{equation}
 d^* = \frac{1}{n_d}\Bigg(\sum_{i\in N^d}d_i + T^*\Bigg).
 \end{equation}
\end{corollary}

In a large platform with free entry, the cutoff $d^*$ approaches zero because the ratio of total fake streams to real streams is bounded. Thus, only artists with negligible real streams, who would otherwise earn almost nothing, choose to engage in fraud. At the same time, the ratio of fake streams to real streams may still approach the upper bound $V-1$.

To summarize, when fraud technology is strong, cheating is confined to low-streamshare artists and aggregate fake streams are bounded. In a large platform with free entry, the worst fraud equilibrium does not occur; instead, only artists with negligible royalties engage in fraud.

\section{Fairness comparison}\label{sec:fairness}
So far we have studied equilibrium outcomes under weak and strong fraud technologies. The fraud-free equilibrium is efficient but requires sufficiently weak fraud technology, while the fraud equilibrium features bounded social waste. The efficiency comparison is therefore clear: strong fraud technology leads to an inefficient outcome. 

However, efficiency is not the only criterion.
From the perspective of fairness, we show that fraud may sometimes improve the distribution of payoffs. The fairness criterion considered here is egalitarianism: social welfare depends on the agent with the lowest utility.

\begin{definition}
 Fix $N$, and let $u$ and $v$ denote two utility profiles. We say that $u$ is fairer than $v$ if $\min_{i\in N} u_i > \min_{i\in N} v_i$.
\end{definition}

We compare the utility profile in the fraud equilibrium with that in the fraud-free situation, where utilities depend only on artists' real streamshares.

\begin{proposition}\label{prop:index2}
 In a fraud equilibrium, artists' utilities have the same ranking as in the fraud-free situation and therefore coincide with the ranking of real streamshares. The utility profile in a fraud equilibrium is fairer than that in the fraud-free situation if and only if
 \begin{equation*}
 1 + T^* < V - \sqrt{d_{\min}V(V-1)}.
 \end{equation*}
\end{proposition}

In the fraud equilibrium, the utilities of honest artists decrease because their revenues depend only on real streamshare, while total streams increase due to fraud. Fraud redistributes payoffs from high-streamshare artists to low-streamshare artists by diluting stream value. This redistribution can raise the minimum utility and thereby improve egalitarian fairness. Combining Proposition \ref{prop:index2} with Proposition \ref{prop:index1} yields an equivalent condition for when the fraud equilibrium is fairer.

\begin{corollary} \label{coro:index4}
 The worst fraud equilibrium is always fairer than the fraud-free situation if and only if
 \begin{equation*}
     d_{\min} < \frac{1}{n^2}\frac{V}{V-1}.
 \end{equation*}
\end{corollary}

In particular, with free entry ($d_{\min}=0$), the fraud equilibrium is always fairer than the fraud-free situation under this criterion.

\section{Discussion on alternative methods}\label{sec:weighted}
\subsection{Parametric weighted rule}
From the analysis above, we acknowledge that efficiency would be lost under a strong fraud technology. To address this issue and restore a fraud-free equilibrium through alternative methods, it is not necessary to abandon pro-rata entirely in favor of user-centric as suggested by \cite{dimont2018}. Instead, we propose a parametric weighted rule to share the revenue.

The rule is a convex combination of pro-rata and user-centric, designed to balance the platform's preference for pro-rata with its limited ability to detect fraud. The revenue paid to artist $i$ under the weighted rule with parameter $\alpha\in(0,1]$ is
\begin{equation*}
 R_i^{pw}(\alpha) = \alpha R_i^{pr} + (1-\alpha) R_i^{uc},
\end{equation*}
where a higher $\alpha$ moves the rule closer to pro-rata. The corresponding net gain is
\begin{equation*}
 G_i^{pw}(x; \alpha) =
 \frac{m \overline{\lambda} d_i + \lambda_0 x_i}{m\overline{\lambda} + \lambda_0\sum_{j\in N}x_j}
 \Bigg(m + \sum_{j\in N}x_j\Bigg)\beta \alpha
 + \Bigg(\sum_{k\in M}\pi_{ik} + x_i\Bigg)\beta (1-\alpha)
 - \delta x_i.
\end{equation*}
Consider the same game with strategy vector $t$, and define the utility of artist $i\in N$ as
$u_i^{pw}(t)=\frac{\lambda_0 G_i^{pw}(x)}{m\overline{\lambda}(\delta-\beta)}$. Then
\begin{equation}
 u_i^{pw}(t;\alpha) =
 \frac{d_i + t_i}{1 + \sum_{j\in N}t_j}V\alpha
 - t_i
 + \frac{1}{\xi}\Bigg[d_i\alpha
 + \frac{\lambda_0}{m\overline{\lambda}}\sum_{k\in M}\pi_{ik}(1-\alpha)\Bigg],
\end{equation}
where the last term does not depend on artist $i$'s choice of $t_i$.

Repeating the earlier analysis under this rule yields a generalization of Theorem \ref{thm:index2}.

\begin{theorem}\label{thm:index3}
 Under the weighted rule with parameter $\alpha$, $t^* = \mathbf{0}$ is the strictly dominant strategy of all artists if and only if
 $\lambda_0 \leq \overline{\lambda}[1+\frac{\xi}{\alpha(1-d_{\min})}]$.
\end{theorem}

Theorem \ref{thm:index3} shows that the threshold for fraud-proofness increases as $\alpha$ decreases. When $\alpha=1$, the condition reduces to the original pro-rata threshold. As $\alpha\to 0$, the threshold approaches infinity, reflecting the fact that user-centric is inherently immune to fraud.

Theorem \ref{thm:index3} also allows us to address the dual question: given a strong fraud technology $\lambda_0 > \overline{\lambda}(1 + \frac{\xi}{1-d_{\min}})$, what values of $\alpha$ guarantee efficiency? The following proposition provides an answer.

\begin{proposition}
 To guarantee a fraud-free equilibrium under a strong fraud technology, the parameter $\alpha$ must satisfy
 \begin{equation*}
 \alpha \leq \frac{1}{(1-d_{\min})V}.
 \end{equation*}
\end{proposition}

This condition provides a practical guideline: platforms can mitigate fraud risk by reducing $\alpha$, thereby shifting the rule toward user-centric. The required adjustment depends on $V$, which relates to $\lambda_0$, $\overline{\lambda}$, $\delta$, and $\beta$, and on the lowest streamshare $d_{\min}$. In this sense, the proposition identifies a family of weighted rules that preserve efficiency under technological constraints.

\subsection{Spotify's modernized royalty system}
At the end of 2023, Spotify announced several new policies to ``better support those most dependent on streaming revenues as part of their livelihood.''\footnote{The policies include charging for artificial streaming, tracking monetization eligibility, and regulating noise recordings. For more details, see \href{https://artists.spotify.com/en/blog/modernizing-our-royalty-system}{https://artists.spotify.com/en/blog/modernizing-our-royalty-system}}
The qualification policy is our focus: it sets a stream threshold that a track must meet to be eligible to receive royalties. In particular, only eligible tracks, those reached at least 1,000 streams in the past year, enter the pro-rata denominator.

The policy nonetheless affects fraud incentives. Adapting the policy to our model, let $\widehat{\lambda}$ denote the qualification threshold of streams, and let $\widehat{d}=\frac{\widehat{\lambda}}{m\overline{\lambda}}$ denote the corresponding streamshare threshold. We treat each track as an artist for simplicity. An artist $i\in N$ shares revenue if $d_i+t_i\ge \widehat{d}$; otherwise, the artist pays the fraud cost and receives zero from the revenue pool. Hence any strategy $t_i\in(0,\widehat{d}-d_i)$ is strictly dominated by 0 for artists with $d_i<\widehat{d}$, since it incurs cost without qualifying for payment.

We argue that this policy may strictly increase fake streams and harm all artists. To illustrate this possibility, we focus on a reasonable range of thresholds $\widehat{d}$ that are not too high: an overly demanding threshold could affect artists who rely on streaming revenue, and a full characterization for arbitrary thresholds is beyond the scope of this paper. Within the range we consider, dishonest artists in the baseline equilibrium still cheat.

Consider a value of $V$ for which the baseline model yields a fraud equilibrium $t^*$ with a set of dishonest artists $N^d$ and a cutoff $d^*$. If the platform chooses a low qualification threshold $\widehat{d}\le d^*$, the policy fails: dishonest artist $i\in N^d$ with $d_i<\widehat{d}\le d^*$ still choose $t_i^*=d^*-d_i\ge \widehat{d}-d_i$, which is unaffected by eliminating dominated strategies.

Suppose instead that the platform chooses a threshold $\widehat{d}$ slightly above $d^*$.\footnote{The range of slightly high $\widehat{d}$ is non-empty, as shown in Appendix \ref{apx:index15}. In this range, all artists cheat at least the same as in the baseline equilibrium, and no artist is eliminated.} Then a low-streamshare artist chooses either $t_i=0$ or $t_i\ge \widehat{d}-d_i$. In a new equilibrium where all initial cheaters still cheat, their best response is the just-qualify strategy $\widehat{d}-d_i$, because the resulting increase in total fake streamshare reduces the marginal return to additional fraud. Let $\widehat{t}$ be defined by $\widehat{t}_i=\max\{\widehat{d}-d_i,0\}$ for all $i\in N$, and $\widehat{T}$ denote aggregate fraud under $\widehat{t}$. The strategy profile $\widehat{t}$ is the equilibrium under threshold $\widehat{d}$. In this equilibrium, artist $i$ with $\widehat{d}>d_i$ will cheat more, i.e., $\widehat{t}_i>t_i^*$; and artist $i$ with $\widehat{d}\leq d_i$ will keep honest, i.e., $\widehat{t}_i = t^*$, and thus $\widehat{T}-\widehat{t}_i>T^*-t_i^*$. Consequently, for all $i\in N$,
\begin{equation*}
 u_i(\widehat{t})
 = \frac{d_i + \widehat{t}_i}{1+\widehat{T}}V - \widehat{t}_i + \frac{d_i}{\xi}
 \le \frac{d_i + \widehat{t}_i}{1+\widehat{t}_i + T^* - t_i^*}V - \widehat{t}_i + \frac{d_i}{\xi}
 = u_i(\widehat{t}_i, t_{-i}^*)
 \le u_i(t^*),
\end{equation*}
where two inequalities cannot both hold with equality, so at least one is strict. Thus all artists are worse off. In addition, it reduces both efficiency and fairness.

\section{Concluding remarks}\label{sec:conclusion}

This paper develops a tractable model of strategic manipulation by artists on music streaming platforms under the pro-rata revenue sharing rule. The framework is flexible enough to accommodate evolving detection methods and increasingly sophisticated fraud strategies, including those enabled by machine learning and advanced identification technologies. It shows that pro-rata can be robust to fraud when fraud technology is weak, yielding an efficient fraud-free equilibrium. When fraud technology is strong, a unique fraud equilibrium exists, but aggregate fake streams remain bounded because increased total streams dilute the value of manipulation. Fraud may also compress payoffs and improve egalitarian fairness despite efficiency losses. Therefore, this paper defends the pro-rata revenue sharing rule against concerns about click-fraud.

\bibliography{main}

\newpage

\appendix
\section{Proofs}\label{apx:proofs}

\subsection{Proof of Lemma \ref{lem:index2}} \label{apx:index2}
Let $\tau_i = \sum_{j\in N\setminus \{i\}} t_j$ denote the summation of all agents' strategy except agent $i$. Rewrite the utility function as $v_i(t_i, \tau_i)$. That is,
\begin{equation}
    v_i(t_i, \tau_i) = \frac{d_i + t_i}{1 + t_i + \tau_i}V - t_i + \frac{d_i}{\xi}. \label{eq:ind22}
\end{equation}
To show $t_i = 0$ is a strictly dominant strategy of agent $i$, one needs to show that for all $\tau_i \geq 0$, it holds for all $t_i > 0$ that $v_i(0,\tau_i) > v_i(t_i, \tau_i)$. It is
\begin{equation}
    \bigg(\frac{d_i}{1 + \tau_i} - \frac{d_i+t_i}{1 + t_i+\tau_i}\bigg)V + t_i > 0,\label{eq:ind30}
\end{equation}
which is equivalent to
\begin{equation}
    (1+\tau_i)t_i + (1 + \tau_i)^2 + (d_i-1-\tau_i)V > 0. \label{eq:ind24}
\end{equation}
So the statement is equivalent to that for all $\tau_i\geq0$,
\begin{equation*}
    (1 + \tau_i)^2 + (d_i-1-\tau_i)V \geq 0, 
\end{equation*}
which is equivalent to 
\begin{equation}
    \bigg(\tau_i + 1 - \frac{1}{2}V\bigg)^2 \geq \frac{1}{4}V^2 - d_iV. \label{eq:ind21}
\end{equation}

Case 1: $V\leq 2$. (\ref{eq:ind21}) holds for all $\tau_i\geq 0$ would be equivalent to
\begin{equation*}
     \bigg(1 - \frac{1}{2}V\bigg)^2 \geq \frac{1}{4}V^2 - d_iV. 
\end{equation*}
It is
\begin{equation*}
    V \leq \frac{1}{1-d_i}.
\end{equation*}
If $d_i \leq \frac{1}{2}$, there is $\frac{1}{1-d_i}\leq 2$. So $V\leq \frac{1}{1-d_i}$. If $d_i > \frac{1}{2}$, there is $\frac{1}{1-d_i} > 2$. So $V\leq 2$.

Case 2: $V > 2$. (\ref{eq:ind21}) holds for all $\tau_i\geq 0$ would be equivalent to
\begin{equation*}
    0 \geq \frac{1}{4}V^2 - d_iV.
\end{equation*}
That is,
\begin{equation*}
    V\leq 4d_i.
\end{equation*}
If $d_i \leq \frac{1}{2}$, there is $V \leq 4d_i \leq 2 < V$, which is not the case. If $d_i > \frac{1}{2}$, there is $2 < V \leq 4d_i$. So $V\leq 4d_i$.

To sum up, if $d_i \leq \frac{1}{2}$, $V \leq \frac{1}{1-d_i}$; if $d_i > \frac{1}{2}$, $V\leq 4d_i$. 
\qed

\subsection{Proof of Lemma \ref{lem:index12}}\label{apx:index6}
To show Lemma \ref{lem:index12}, three steps are constructed. First step shows the existence of Nash equilibrium for $V > 0$. Second step shows the uniqueness of Nash equilibrium for $V \geq 1$. Third step shows that $t^*=\mathrm{0}$ is not a Nash equilibrium when $V > \frac{1}{1- d_{\min}}$.

Step 1 (Existence): Assume $V > 0$. The derivative of agent $i$'s utility function (\ref{eq:ind22}) is
\begin{equation}
    \frac{\partial v_i(t_i, \tau_i)}{\partial t_i} = \frac{1-d_i+\tau_i}{(1+t_i+\tau_i)^2}V-1, \label{eq:ind23}
\end{equation}
which approaches $-1$ as $t_i\rightarrow+\infty$. The second-order derivative is
\begin{equation*}
    \frac{\partial^2 v_i(t_i, \tau_i)}{\partial t_i^2} = -2 \frac{(1-d_i+\tau_i)}{(1+t_i+\tau_i)^3}V,
\end{equation*}
which is negative. So the utility function is strictly concave. 

Now prove the following claim: There exists an upper bound $\overline{t}_i \geq 0$ such that for all $\tau_i\geq 0$, $\frac{\partial v_i(t_i, \tau_i)}{\partial t_i} < 0$ holds for all $t_i > \overline{t}_i$. Let $\overline{t}_i = \frac{1}{2}V$, then it is to show that for all $\tau_i\geq 0$, all $t_i > \frac{V}{2}$, 
\begin{equation*}
    (1-d_i+\tau_i)V < (1+t_i+\tau_i)^2. 
\end{equation*}
It is equivalent to that for all $\tau_i\geq 0$,
\begin{equation*}
    (1-d_i + \tau_i)V \leq \bigg(1 + \frac{1}{2}V+\tau_i\bigg)^2.
\end{equation*}
That is, 
\begin{equation*}
    (1 + \tau_i)^2 + \frac{1}{4}V^2 + d_iV \geq 0,
\end{equation*}
which holds for all $\tau_i \geq 0$.

So all strategies exceeding $\frac{1}{2}V$ are strictly dominated. Therefore the strategy set of agent $i$ can be restricted into $[0, \frac{1}{2}V]$, which is compact and convex. Combining with the continuity and convexity of utility function, Nash theorem ensures the existence of Nash equilibrium.

Step 2 (Uniqueness): Take $V\geq 1$. Let $t^*$ denote the Nash equilibrium and $T^* = \sum_{i\in N}t^*_i$ denote the aggregate strategy in equilibrium. Existence of Nash equilibrium implies that there exists a set of agents $N^d\subseteq N$ with positive strategy in equilibrium. Due to concavity of utility function, for all $i\in N^d$, there are $t^*_i > 0$ and $\frac{\partial v_i(t^*_i, \tau^*_i)}{\partial t_i} = 0$, which, from (\ref{eq:ind23}), is equivalent to 
\begin{equation}
    d_i + t^*_i = 1 + T^* - \frac{1}{V}(1 + T^*)^2. \label{eq:ind34}
\end{equation}
Also for all $j\in N\setminus N^d$, there are $t^*_j=0$ and 
\begin{equation}
    d_j \geq  1 + T^* - \frac{1}{V}(1 + T^*)^2.\label{eq:ind35}
\end{equation}
So the positivity of $t^*_i$ is determined by $1 + T^* - \frac{1}{V}(1 + T^*)^2$ and $d_i$. 

Consider a helper function $f_i(T^*) = \frac{d_i + t^*_i}{1+T^*}$ for all $i\in N$, of which the form depends on $d_i$.

Case 1: $d_i \geq \frac{1}{4}V$. The following inequality always holds:
\begin{equation*}
    \bigg(1+T^*-\frac{1}{2}V\bigg)^2 \geq \frac{1}{4}V^2 - d_i V,
\end{equation*}
which is equivalent to (\ref{eq:ind35}). So in equilibrium agent $i$ would always take strategy $t^*_i = 0$. Thus $f_i(T^*) = \frac{d_i}{1+T^*}$, which is continuous and strictly decreasing on $\mathbb{R}_+$.

Case 2: $d_i < \frac{1}{4}V$. From (\ref{eq:ind35}), we know that $t^*_i = 0$ if either $1+T^* \leq \frac{1}{2}V - \sqrt{\frac{1}{4}V^2 - d_i V}$, or $1+T^* \geq \frac{1}{2}V + \sqrt{\frac{1}{4}V^2 - d_i V}$. Otherwise, $d_i + t^*_i = 1 + T^* - \frac{1}{V}(1 + T^*)^2$. So that 
\begin{equation*}
    f_i(T^*) = \left\{
    \begin{array}{ll} 
        \frac{d_i}{1 + T^*} & \text{if $T^* \leq \frac{1}{2}V - \sqrt{\frac{1}{4}V^2 - d_iV}-1$;} \\
        1 - \frac{1}{V}(1 + T^*) & \text{if $\frac{1}{2}V - \sqrt{\frac{1}{4}V^2 - d_iV}-1 < T^* < \frac{1}{2}V + \sqrt{\frac{1}{4}V^2 - d_iV}-1$;} \\
        \frac{d_i}{1 + T^*} & \text{if $T^* \geq \frac{1}{2}V + \sqrt{\frac{1}{4}V^2 - d_iV}-1$,} 
    \end{array}\right.
\end{equation*}
which is continuous and strictly decreasing on $\mathbb{R}_+$. 

Let $F(T^*) \equiv \sum_{i\in N} f_i(T^*)$. Since $f_i(T^*)$ is continuous and strictly decreasing on $\mathbb{R}_+$ for all $i\in N$, $F(T^*)$ is continuous and strictly decreasing on $\mathbb{R}_+$. 
Notice that 
\begin{equation*}
    1 = \frac{1 + T^*}{1 + T^*} = \sum_{i\in N} \frac{d_i + t^*_i}{1 + T^*} = F(T^*).
\end{equation*}
From $f_i(T^*) = \frac{d_i + t^*_i}{1 + T^*} \geq \frac{d_i}{1+T^*}$, there is $F(T^*)\geq \sum_{i\in N}\frac{d_i}{1+T^*} = \frac{1}{1+T^*}$. So $F(0) \geq 1$. Combining with $F(V-1) = \frac{1}{V} \leq 1$, as well as the constant LHS, it yields a unique aggregate strategy $T^*\in [0, V-1]$. Therefore the uniqueness can be derived by unique $T^*$ as well as constant $d_i$ for all $i\in N$. 

Step 3 (Non-existence at zero): This step shows that $t^*=\mathbf{0}$ is not a Nash equilibrium when $V > \frac{1}{1-d_{\min}}$ by contradiction. Suppose $t^*=\mathbf{0}$ is a Nash equilibrium. Consider the pivotal agent $i$ with $d_i = d_{\min}$. There is $v_i(0, 0) \geq v_i(t_i, 0)$ for all $t_i > 0$. Substituting $\tau_i =0$ into (\ref{eq:ind24}), it is, 
\begin{equation}
    t_i + 1 \geq (1-d_{\min})V. \label{eq:ind25}
\end{equation}
One can find a strategy
\begin{equation*}
    \widetilde{t}_i = \sqrt{(1-d_{\min})V} - 1 > 0,
\end{equation*}
as $V > \frac{1}{1-d_{\min}} \geq 1$.
Substituting $\widetilde{t}_i$ into (\ref{eq:ind25}), there is
\begin{equation*}
    \sqrt{(1-d_{\min})V} \geq (1-d_{\min})V,
\end{equation*}
which is equivalent to $(1-d_{\min})V\leq 1$ contradicting $V> \frac{1}{1-d_{\min}}$. Thus $t^*=\mathbf{0}$ is not a Nash equilibrium.

To sum up, if $V > \frac{1}{1-d_{\min}}$, there exists a unique fraud equilibrium, where at least one agent takes positive strategy. 
\qed

\subsection{Proof of Theorem \ref{thm:index4}}\label{apx:index11}
From Lemma \ref{lem:index12}, there exists a unique fraud equilibrium. Thus there is a unique non-empty subset of agents $N^d\subseteq N$ taking positive strategies in equilibrium. Following lemma shows that the agents can be separated according to real streamshare $d_i$ and those who cheat are those with less real streamshare.

\begin{lemma}\label{lem:index93}
    Suppose $t^*$ is a fraud equilibrium, and $N^d$ is a set of agents who take positive strategies in equilibrium. Then for all pair $i,j\in N$: 
    \begin{enumerate}[label=(\alph*)]
        \item If $i\in N^d$ and $d_j\leq d_i$, then $j\in N^d$;
        \item If $i\in N\setminus N^d$ and $d_j\geq d_i$, then $j\in N\setminus N^d$.
    \end{enumerate}
\end{lemma}
\noindent\textit{Proof}. (a) Given $i\in N^d$ and $d_j\leq d_i$, suppose $j\in N\setminus N^d$. There is $t^*_i > 0$, $t^*_j= 0$, $\frac{\partial v_i(t_i^*, \tau_i^*)}{\partial t_i}= 0$, and $\frac{\partial v_j(t_j^*, \tau_j^*)}{\partial t_j}\leq 0$. From (\ref{eq:ind34}) and (\ref{eq:ind35}), it implies
\begin{equation*}
    1 + T^* - \frac{1}{V}(1 + T^*)^2 \leq d_j + t^*_j = d_j \leq d_i < d_i + t^*_i = 1 + T^* - \frac{1}{V}(1 + T^*)^2,
\end{equation*}
which is a contradiction. Thus $j\in N^d$.

(b) Conversely, given $i\in N\setminus N^d$ and $d_j\geq d_i$, suppose $j\in N^d$. There is $t^*_i = 0$, $t^*_j > 0$, $\frac{\partial v_i(t_i^*, \tau_i^*)}{\partial t_i} \leq 0$ and $\frac{\partial v_j(t_j^*, \tau_j^*)}{\partial t_j}= 0$. From (\ref{eq:ind34}) and (\ref{eq:ind35}), it implies
\begin{equation*}
    1 + T^* - \frac{1}{V}(1 + T^*)^2 = d_j + t^*_j > d_j \geq d_i = d_i + t^*_i \geq 1 + T^* - \frac{1}{V}(1 + T^*)^2, 
\end{equation*}
which is a contradiction. Thus $j\in N\setminus N^d$.
\qed

Next study the fraud equilibrium with set $N^d$. For all $i,j\in N^d$, repeating (\ref{eq:ind34}) gives
\begin{equation*}
    d_i + t_i^* = 1 + T^* - \frac{1}{V}(1 + T^*)^2 = d_j + t^*_j.
\end{equation*}
Notice that $T^* = \sum_{i\in N^d} t^*_i$. Summing up (\ref{eq:ind34}) for all $i\in N^d$, it is
\begin{equation}
    \sum_{i\in N^d} d_i + T^* = n_d \bigg[ (1+T^*)-\frac{1}{V}(1+T^*)^2 \bigg]. \label{eq:ind27}
\end{equation}
It is equivalent to
\begin{equation*}
    \bigg(1+T^* - \frac{n_d-1}{2n_d}V\bigg)^2 = \bigg(\frac{n_d-1}{2n_d}V\bigg)^2 + \frac{1}{n_d}\Bigg(1-\sum_{i\in N^d}d_i\Bigg)V,
\end{equation*}
where the right-hand side (RHS) is non-negative.
Thus
\begin{equation*}
    1+T^* = \sqrt{\bigg(\frac{n_d-1}{2n_d}V\bigg)^2 + \frac{1}{n_d}\Bigg(1-\sum_{i\in N^d}d_i\Bigg)V} + \frac{n_d-1}{2n_d}V.
\end{equation*}
\qed

\subsection{Proof of Proposition \ref{prop:index1}}\label{apx:index13}

Let $r(N^d) = 1+T^*$ a function of set of agents $N^d$ with positive strategies in equilibrium. That is,
\begin{equation*}
    r(N^d) = \sqrt{\bigg(\frac{n_d-1}{2n_d}V\bigg)^2 + \frac{1}{n_d}\Bigg(1-\sum_{i\in N^d}d_i\Bigg)V} + \frac{n_d-1}{2n_d}V.
\end{equation*}
Consider two helper functions:
\begin{align*}
    \underline{r}(s) &= \frac{s-1}{s}V; \\ 
    \overline{r}(s) &= \sqrt{\bigg(\frac{s-1}{2s}V\bigg)^2 + \frac{1}{s}V} + \frac{s-1}{2s}V,\label{eq:ind31}
\end{align*}
where $s$ is treated as a continuous variable. Notice that for all possible $N^d$ in fraud equilibrium, there is
\begin{equation}
    \underline{r}(n_d) \leq r(N^d)\leq \overline{r}(n_d).
\end{equation}
It's obvious that $\underline{r}(s)$ is strictly increasing in $s$. Notice that $\overline{r}$ is a positive root of the following equation:
\begin{equation*}
    \overline{r}^2 - \frac{s-1}{s}V\overline{r} - \frac{1}{s}V = 0,
\end{equation*}
while another root is negative. Let $F(\overline{r}, s) = \overline{r}^2 - \frac{s-1}{s}V\overline{r} - \frac{1}{s}V$. Notice that $F(1, s) = 1-V < 0$, and $F(\overline{r}, s) \rightarrow+\infty\ (\overline{r}\rightarrow+\infty)$. So that $\overline{r} > 1$. The derivative of $\overline{r}(s)$ is
\begin{equation*}
\begin{split}
    \overline{r}'(s) 
    &= \frac{1}{2s^2}V\Bigg[\frac{\frac{s-1}{2s}V-1}{\sqrt{(\frac{s-1}{2s}V)^2 + \frac{1}{s}V}} + 1 \Bigg]. \\
    &= \frac{1}{2s^2}V \frac{\sqrt{(\frac{s-1}{2s}V)^2 + \frac{1}{s}V} + \frac{s-1}{2s}V-1}{\sqrt{(\frac{s-1}{2s}V)^2 + \frac{1}{s}V}} \\
    &= \frac{1}{2s^2}V \frac{\overline{r}(s)-1}{\sqrt{(\frac{s-1}{2s}V)^2 + \frac{1}{s}V}},
\end{split}
\end{equation*}
which is positive. So $\overline{r}(s)$ is strictly increasing. Also,
\begin{equation*}
    \overline{r}(n) = \sqrt{\bigg(\frac{n-1}{2n}V\bigg)^2 + \frac{1}{n}V} + \frac{n-1}{2n}V,
\end{equation*}
which approaches its upper bound $V$ as $n\rightarrow+\infty$. Combining with the fact that $\underline{r}(n)$ also approaches $V$ as well. The upper bound of $1+T^*$ is thus $V$.
\qed

\subsection{Proofs of Corollaries \ref{coro:index1}, \ref{coro:index2}, and \ref{coro:index4}}\label{apx:index12}

\noindent\textit{Proof of Corollary \ref{coro:index1}}. From statement (a) of Theorem \ref{thm:index4}, equilibrium strategy of agent $i\in N^d$ is decreasing in $d_i$. All agents taking positive strategies is thus equivalent to that the agent with $d_{\mathrm{max}}$ takes positive strategy. Suppose agent $i$ is the agent with $d_i = d_{\mathrm{max}}$. From (\ref{eq:ind34}), the equilibrium strategy of agent $i$ is
\begin{equation*}
    d_{\mathrm{max}} + t_i^* = 1+T^* - \frac{1}{V}(1+T^*)^2.
\end{equation*}
Then $t_i^* > 0$ is equivalent to
\begin{equation*}
    d_{\mathrm{max}} < 1+T^* - \frac{1}{V}(1+T^*)^2.
\end{equation*}
Notice that $1+T^* = \frac{n-1}{n}V$ holds in the worst fraud equilibrium. So it is equivalent to
\begin{equation*}
    d_{\mathrm{max}} < \frac{n-1}{n}V - \frac{1}{V} (\frac{n-1}{n}V)^2 = \frac{n-1}{n^2}V.
\end{equation*}
Thus the equivalent condition is obtained.
$\qedsymbol$

\noindent\textit{Proof of Corollary \ref{coro:index2}}. One can find a threshold
\begin{equation}
    d^* = 1+T^* - \frac{1}{V}(1+T^*)^2. \label{eq:ind28} 
\end{equation}
Now check two cases. For all $i\in N^d$ with $t_i^* > 0$. From (\ref{eq:ind34}), there is
\begin{equation*}
    1+T^* - \frac{1}{V}(1+T^*)^2 = d_i +t^*_i > d_i,
\end{equation*}
which is equivalent to $d_i < d^*$. Also, for all $j\in N\setminus N^d$ with $t^*_j = 0$. From $\frac{\partial v_j(t_j^*, \tau_j^*)}{\partial t_j}\leq 0$, there is
\begin{equation*}
    1+T^* - \frac{1}{V}(1+T^*)^2 \leq d_j + t^*_j = d_j,
\end{equation*}
which is exactly $d_j \geq d^*$. Substituting (\ref{eq:ind27}) into (\ref{eq:ind28}), formula shown in corollary can be derived.
\qed

\noindent\textit{Proof of Corollary \ref{coro:index4}}.
From Theorem \ref{thm:index4}, as well as $\sum_{i\in N}d_i = 1$, there is $1+T^* = \frac{n-1}{n}V$. 
Combining with Proposition \ref{prop:index2}, the statement holds if and only if 
\begin{equation*}
    \frac{n-1}{n}V < V - \sqrt{d_{\min}V(V-1)},
\end{equation*}
which is equivalent to the formula. 
\qed

\subsection{Proof of Proposition \ref{prop:index2}}\label{apx:index14}
Consider two utility profiles $u^*$ and $u^0$, corresponding to that in fraud equilibrium and the fraud-free situation. Obviously, $u_i^0 = u_i(\mathbf{0}) = d_i(\frac{1}{\xi}+V)$ for all $i\in N$.
From (\ref{eq:ind34}) and (\ref{eq:ind28}), for all agent $i\in N^d$ with positive equilibrium strategy, there is
\begin{equation*}
    d_j+t_j^* = d^*,
\end{equation*}
while for all agent $j\in N\setminus N^d$ with $t_j^* = 0$, there is $d_j\geq d^*$. So that for all agent $i\in N^d$, 
\begin{equation}
    u_i^* = \frac{d^*}{1+T^*}V-d^* + \bigg(1 + \frac{1}{\xi}\bigg)d_i. \label{eq:ind29}
\end{equation}
The first two terms are independent of $i\in N^d$. Thus for all agents in $N^d$, utilities among profile $u^*$ follow the same order as those in $u^0$. 

Next we show that, in partial fraud equilibrium, the utility of any agent in $N\setminus N^d$ would be higher than that of all agents in $N^d$. Consider any pair $i\in N^d$ and $j\in N\setminus N^d$ in partial fraud equilibrium, where $N\setminus N^d$ is not empty. Notice that
\begin{equation*}
    u_j^* = d_j\bigg(\frac{1}{\xi}+\frac{V}{1+T^*}\bigg) = \frac{d_j}{1+T^*}V - d^* + d^* +\frac{1}{\xi}d_j > \frac{d^*}{1+T^*}V - d^* + d_i + \frac{1}{\xi}d_i = u_i^*,
\end{equation*}
where the inequality is obtained from $d_j \geq d^*>d_i$. Thus we can conclude that utilities in the fraud equilibrium follow the same order as those in the fraud-free situation.

To show the fairness property, the comparison between two profiles for the utility of pivotal agent with $d_{\min}$ is needed. Denote pivotal agent's utility in fraud equilibrium and that in fraud-free situation by $u^*_{\min}$ and $u^0_{\min}$, respectively. Combining (\ref{eq:ind29}) and $u_i^0 = d_i(\frac{1}{\xi}+V)$, there is
\begin{equation*}
\begin{split}
    u^*_{\min} - u^0_{\min} 
    &= \frac{d^*}{1+T^*}V-d^* + \bigg(1+\frac{1}{\xi}\bigg)d_{\min} - d_{\min}\bigg(\frac{1}{\xi}+V\bigg) \\
    &= \frac{1}{V}[(1+T^*)- V]^2 - d_{\min}(V-1).
\end{split}
\end{equation*}
So that the fraud equilibrium is fairer, i.e., $u^*_{\min} > u^0_{\min}$, is equivalent to
\begin{equation*}
    [(1+T^*)- V]^2 > d_{\min}V(V-1).
\end{equation*}
That is, either
$1+T^* > V + \sqrt{d_{\min}V(V-1)}$, or 
$1+T^* < V - \sqrt{d_{\min}V(V-1)}$. From Proposition \ref{prop:index1}, there is $1+T^* \leq V$. Thus only the later range is kept. 
\qed

\subsection{Proof of Theorem \ref{thm:index3}}\label{apx:index7}
Similar to the proof of Lemma \ref{lem:index2}. Rewrite the parametric utility function of agent $i$ as follows.
\begin{equation}
    v_i(t_i, \tau_i; \alpha) = \frac{d_i + t_i}{1 + t_i + \tau_i}V\alpha - t_i + \frac{1}{\xi}\Bigg[d_i\alpha + \frac{\lambda_0}{m\overline{\lambda}}\sum_{k\in M}\pi_{ik}(1-\alpha)\Bigg]. 
\end{equation}
$t_i=0$ is a strictly dominant strategy of agent $i$ is equivalent to that for all $\tau_i \geq 0$, $v_i(0, \tau_i;\alpha) > v_i(t_i, \tau_i;\alpha)$ holds for all $t_i > 0$. That is,
\begin{equation*}
    \bigg(\frac{d_i}{1 + \tau_i} - \frac{d_i+t_i}{1 + t_i+\tau_i}\bigg)V\alpha + t_i > 0,
\end{equation*}
which is exactly (\ref{eq:ind30}) if $V$ is replaced with $V\alpha$. Repeating the analysis of Lemma \ref{lem:index2} gives the equivalent formula. That is, $V\alpha\leq \frac{1}{1-d_i}$ if $d_i \leq \frac{1}{2}$, and $V\alpha\leq 4d_i$ if $d_i > \frac{1}{2}$. 
Thus the theorem can be immediately derived.
\qed

\subsection{Existence of slightly high threshold}\label{apx:index15}

Now verify that there exists such policy $\widehat{d}$ leading to a equilibrium that all artists are not eliminated. The focus is on the pivotal artist $i$, who is most likely to switch from just-qualifying to honesty as $\widehat{d}$ increases. The pivotal artist will stay and play the just-qualify strategy when
$u_i(\widehat{t}) \geq 0$. 
Notice that $1+T^*<1+\widehat{T}\leq 1+T^*+n(\widehat{d}-d^*)$. There is
\begin{equation*}
\begin{split}
    u_i(\widehat{t}) 
    &= \frac{\widehat{d}}{1+\widehat{T}}V - \widehat{d} + \bigg(1+\frac{1}{\xi}\bigg)d_{\min} \\
    &\geq \frac{d^* + (\widehat{d}-d^*)}{1+T^* + n(\widehat{d}-d^*)}V  -d^* - (\widehat{d}-d^*) + \bigg(1+\frac{1}{\xi}\bigg)d_{\min}.
\end{split}
\end{equation*} 
Let $\theta \equiv \widehat{d}-d^* > 0$ and consider the following helper function of the RHS.
\begin{equation*}
    H(\theta) = \frac{d^* + \theta}{1+T^* + n\theta}V  -d^* - \theta + \bigg(1+\frac{1}{\xi}\bigg)d_{\min}.
\end{equation*}
Its derivative is
\begin{equation*}
    H'(\theta) = \frac{1+T^*-n d^*}{(1+T^* + n\theta)^2}V - 1.
\end{equation*}
If $\frac{1+T^*}{nd^*} < \frac{V}{V-1}$, then $H'(\theta) < 0$ always hold. Let $\theta^* = 0$ in this case. If $\frac{1+T^*}{nd^*} \geq \frac{V}{V-1}$, let $\theta^* = \max\{\frac{1}{n}[\sqrt{(1+T^*-nd^*)V}-(1+T^*)], 0\}$, then $H'(\theta) < 0$ on $(\theta^*, +\infty)$, and $H'(\theta)>0$ on $(0, \theta^*)$ if $\theta^* > 0$.

From two cases, we have $H(\theta^*)\geq H(0) = u_i(t^*) > u_i(0, t^*_{-i})\geq 0$, and $H(V+(1+\frac{1}{\xi})d_{\min}) = [\frac{d^*+V+(1+\frac{1}{\xi})d_{\min}}{1+T^*+nV + n(1+\frac{1}{\xi})d_{\min}}-1]V - d^* < 0$, where $d^*+V+(1+\frac{1}{\xi})d_{\min}<1+T^*+nV + n(1+\frac{1}{\xi})d_{\min}$. Thus there exists a $\theta^{**}\in (\theta^*, V+(1+\frac{1}{\xi})d_{\min})$ such that $H(\theta^{**})=0$. Combining that $H(\theta)$ is strictly increasing on $(0, \theta^*)$ for positive $\theta^*$, and strictly decreasing on $(\theta^*, V+(1+\frac{1}{\xi})d_{\min})$, it has $H(\theta) \geq 0$ for all $\theta\in [0,\theta^{**}]$. Thus we can find a non-empty range for $\widehat{d}\in (d^*, d^*+\theta^{**}]$, such that
$u_i(\widehat{t})\geq H(\theta^{**})\geq 0$,
verifying the existence of slightly high threshold for the new equilibrium. 
\qed

\end{document}